\documentclass{aipproc}
\layoutstyle{6x9}
\begin{document}

\title{Hayashi and the Thermal Physics of Star-Forming Clouds}


\author{Richard B. Larson}
{address={Yale Astronomy Department, New Haven, CT 06520-8101, USA}}

\maketitle

   A fundamental aspect of star formation theory that was pioneered by Hayashi and his associates, notably Nakano, was the thermal physics of collapsing star-forming clouds.  Early discussions of the theory of star formation had been concerned with the question of whether and under what circumstances an interstellar cloud might be able to collapse under gravity to form a star, and in order to answer that question and study the collapse process it was necessary to know the pressure and therefore the temperature of the gas in the cloud as a function of time and location.  This required understanding the processes that control the temperature in such clouds, and this problem was first addressed systematically in the path-breaking work of Hayashi and Nakano (1965a,b), published in Progress of Theoretical Physics.  

   Hayashi and Nakano considered heating by compression and cosmic rays, cooling by line emission from H$_2$ molecules and heavy ions such as C$^+$, and cooling by thermal emission from dust.  They discussed how a cloud might evolve in a temperature-density diagram, and they concluded that a cloud that began at an arbitrary location in such a diagram would rapidly evolve toward a nearly horizontal locus having a temperature of around 15 to 20K.  If its density was high enough for gravity to overcome pressure, the cloud would then contract along this locus to higher densities until it became opaque to its thermal cooling radiation, at which point the temperature would begin to rise substantially.  I first became aware of this work in late 1965 when a preprint of the paper by Hayashi and Nakano (1965b) was called to my attention by Martin Rees, a young postdoc who had just arrived at Caltech when I was a graduate student beginning work on the protostellar collapse problem under the supervision of Guido M\"unch.  I read the paper of Hayashi and Nakano with great interest, and I was impressed with how ambitious it had been in laying out the whole landscape of the problem, discussing the relevant physical regimes and processes, and taking the first steps toward solving the problem quantitatively with clever approximations, and I realized that I could use some of their results as a starting point for my own work.

   The essential features of the temperature-density diagram of Hayashi and Nakano (1965b) can be seen in a simplified and updated version published a year later by Hayashi in a well-known review article (Hayashi 1966).  The most important feature of this diagram, as noted by Hayashi, is that the temperature-density relation is predicted to be ``nearly horizontal'', or nearly isothermal during the early stages of collapse, the temperature remaining near 10K over many orders of magnitude in density until the optical depth becomes large.  This result was the basis of the widely adopted assumption that the early stages of protostellar collapse are isothermal with a temperature of 10K, which became a standard simplifying assumption in an extensive literature on the theory and simulations of star formation.

   The predicted temperature is not exactly constant, however, but first declines with increasing density until reaching a minimum, after which it rises slowly with increasing density.  As can be seen from Hayashi's diagram, the corresponding Jeans mass initially declines rapidly with increasing density until the temperature reaches its minimum, after which the Jeans mass declines more slowly with increasing density.  The resulting Jeans masses span the entire range of stellar masses, and at the temperature minimum the Jeans mass is a few tenths of a solar mass, about the mass at which the stellar initial mass function peaks (when measured in logarithmic units).  When I made similar calculations and found similar results, I was struck by this correspondence and wondered whether the stellar initial mass function might be determined by the temperature-density relation and whether the Jeans mass at the temperature minimum might represent a preferred mass scale for star formation.  But I did not at the time see how to turn these vague ideas into a theory of the IMF.

   Many years later it had become possible to simulate cloud fragmentation in considerable detail, and a paper by Li, Klessen, and Mac Low (2003) appeared showing that the amount of fragmentation that can occur in a collapsing cloud is indeed quite sensitive to the assumed temperature-density relation.  These authors assumed a polytropic equation of state $P \propto \rho^{\gamma}$ with various values of $\gamma$, and they found that the number of bound fragments that form depends strongly on the value of $\gamma$, decreasing by a large factor when $\gamma$ increases from 0.7 to 1.1, about the range of values relevant for the predicted temperature-density relation (Larson 1985, 2005).  The temperature minimum, where $\gamma = 1$, thus corresponds to a transition from efficient fragmentation when $\gamma < 1$ to relatively limited fragmentation when $\gamma > 1$, and in this sense the Jeans mass at the temperature minimum does indeed represent a preferred mass for star formation, i.e.\ a mass scale below which fragmentation becomes much less probable.

   This strong dependence of the fragmentation efficiency on $\gamma$ can be understood if fragmentation typically occurs in filaments.  This is because for $\gamma < 1$ a cylinder can collapse axially and fragment indefinitely, whereas for $\gamma > 1$ indefinite axial collapse is no longer possible and continuing fragmentation can only occur if the cylinder breaks up into clumps that then collapse more nearly spherically (Larson 2005).  There is indeed now much evidence from both observations and simulations that star formation often involves the fragmentation of filaments.  One of the first theoretical studies showing a clear tendency for gravitational collapse to produce filamentary structure was the paper of Miyama, Narita, and Hayashi (1987); although Hayashi had retired by this time, having retired in 1984, he was still actively working at the frontiers of star formation theory.  Filamentary cloud structure very similar to that predicted by Miyama, Narita, and Hayashi (1987) is seen in observational results like those of Lombardi, Alves, and Lada (2006) for the `Pipe Nebula', a cloud in an early stage of star formation.

   Simulations by Jappsen et al (2005) designed to test the idea that the peak mass of the IMF is determined by the Jeans mass at the temperature minimum show approximate agreement with this prediction, although other factors clearly also play a role and we do not yet have a full understanding of the origin of the stellar IMF.  Nevertheless, temperature-density relations like that discussed above can provide useful insights into how stellar masses and the IMF might vary when parameters such as the metallicity of the star-forming gas are varied.  Low and Lynden-Bell (1976) presented temperature-density relations for a range of metallicities, and their results show that for a metallicity of 0.01 times solar, the temperature-density relation has two distinct minima, one at a lower density associated with line cooling and a second and deeper one at a higher density associated with dust cooling.  The dust-cooling minimum is presumably the more relevant one for the formation of low-mass stars, and when the metallicity is decreased it shifts to higher density and higher temperature roughly along lines of constant Jeans mass, suggesting that the IMF may not vary strongly with metallicity.  Observationally, there is no clear evidence for any dependence of the IMF on metallicity.

   What happens when the metallicity is zero and molecular hydrogen is the only coolant, as is expected for the first stars?  An early paper by Yoneyama (1972) presented a temperature-density relation for H$_2$ cooling that resembled qualitatively the solar-metallicity relation discussed above, with an initial decline in temperature to a minimum followed by a slow rise with increasing density.  The temperature minimum occurs when the level populations of the H$_2$ molecule thermalize, implying that the cooling rate per unit mass stops increasing with increasing density and becomes independent of density.  This is analogous to what happens in the solar-metallicity case discussed above, where the temperature minimum is associated with the transition from line cooling whose rate increases with density to dust cooling whose rate is independent of density.  But the temperature minimum for H$_2$ cooling occurs at a lower density and a much higher temperature of about 200K where the Jeans mass is about 2000 solar masses, suggesting very large masses for the first stars.  Yoneyama (1972), using a somewhat different argument, also predicted very large masses for the first stars.

   In Sato's talk at this meeting, I learned for the first time that in his retirement talk in 1984, Hayashi addressed the formation of the first stars and the thermal properties of metal-free gas.  He mentioned the work of Palla, Salpeter, and Stahler (1983), which showed an initial rise in temperature followed by a decline to a minimum near 300K and finally by a gradual rise to 1000K at the highest densities.  More recent calculations of the temperature-density relation resulting from H$_2$ cooling in a cosmological context by Bromm, Coppi, and Larson (2002) and Abel, Bryan, and Norman (2002) show similar trends, with a strong temperature minimum near 200K followed by a slow rise with increasing density.  Because H$_2$ cooling becomes less efficient at the higher densities where the cooling saturates, the collapse slows down at this point and the gas accumulates in a clump with a mass similar to the Jeans mass at the temperature minimum, or about 500 solar masses, again suggesting very large masses for the first stars.  These investigations followed the collapse of the clump to form a small central accreting core, but they were not able to follow fully the subsequent accretion process and could only estimate final stellar masses between several tens and several hundreds of solar masses.  Yoshida et al (2006) followed the later stages of the collapse in greater detail and found a continuing increase in temperature with density up to about 2000K, estimating a final stellar mass around 100 solar masses.  Recent work has typically shown fragmentation into binary or multiple systems and has favored somewhat smaller final masses of several tens of solar masses; we will hear more at this meeting about the latest developments in this subject.

   Omukai, Hosokawa, and Yoshida (2010) have presented a current temperature-density diagram including all metallicities from solar to zero, and versions of this diagram have been used by these and other authors to discuss the possible implications of different metallicities for stellar masses and the IMF.  For most metallicities, the curves in this diagram show two minima, one at a low density associated with line cooling and a second one at a higher density associated with dust cooling.  The dust-cooling minimum persists to metallicities as low as $10^{-5}$ solar, suggesting that the formation of low-mass stars can occur even at metallicities as low as $10^{-5}$ solar as long as dust is present.  For metallicities of $10^{-6}$ solar or less, however, heavy elements no longer contribute significant cooling and the temperature-density relation converges to that for pure H$_2$ cooling.  The temperature-density relation for H$_2$ cooling has only one prominent temperature minimum at a relatively low density where the Jeans mass is around 500 solar masses, as noted above; at higher densities the temperature rises slowly and relatively smoothly with increasing density, and there are no significant additional minima or other features.  Therefore there appears to be no preferred mass scale smaller than the clump mass of several hundred solar masses, even though the Jeans mass decreases by many orders of magnitude to less than a tenth of a solar mass as the temperature rises slowly at the higher densities, suggesting that the formation of low-mass stars remains possible in principle even for a metallicity of zero.

   However, we cannot draw definite conclusions from such simple temperature-density relations alone.  While such relations may provide valuable insights into what {\it can} happen in a collapsing and fragmenting cloud, they cannot by themselves tell us what actually {\it does} happen; this is because they are based on simplified physics that neglects most dynamical and feedback effects, and star formation is clearly a highly dynamic process.  The only way to determine what actually does happen in a collapsing and fragmenting cloud is to calculate the dynamics in as much detail and with as much realism as possible.  An impressive calculation by Greif et al (2012) that you will hear more about at this meeting has simulated the collapse and fragmentation of a metal-free star-forming clump like those discussed above with ultra-high resolution going all the way down to sub-stellar scales.  As in previous work, a central accreting object forms and begins to grow, and a compact disk then forms around it and fragments into smaller accreting objects whose initial masses are smaller than a tenth of a solar mass.  Thus fragmentation down to the Jeans scale of less than a tenth of a solar mass can indeed occur in such a disk.  But the accreting objects formed in the simulation rapidly grow and begin to merge, which can be understood by noting that at the relevant high density the Jeans length becomes as small as an astronomical unit or less, comparable to the sizes of the accreting protostars.  Merging of the accreting objects must then necessarily occur.  Greif et al (2012) could only follow the initial stages of the accretion and merging process, and the final outcome remains undetermined.  Possibly most of the mass would still end up in one or a small number of massive stars, as anticipated from previous work.  Whether any low-mass stars of zero metallicity survive the merging process and might thus survive to the present time remains to be determined.

   The frontiers of the subject have thus reached a regime where the physics has become much more complicated and challenging and where interstellar and stellar physics merge.  It is thus fitting that this meeting is being held in honor of Hayashi, who made major contributions to both stellar and interstellar physics and who played a major role in launching the modern quantitative study of star formation with rigorous calculations of broad scope that include as much of the relevant physics as possible.

\end{document}